\documentclass[final, conference]{IEEEtran}
\IEEEoverridecommandlockouts
\usepackage{cite}
\usepackage{amsmath,amssymb,amsfonts}
\usepackage{graphicx}
\usepackage{textcomp}
\usepackage[dvipsnames]{xcolor}
\usepackage{braket}
\usepackage{hyperref}
\hypersetup{
    colorlinks=true,
    linkcolor=BurntOrange,
    urlcolor=BrickRed,
    citecolor=Mulberry,
    pdftitle={Automated quantum error mitigation based on probabilistic error reduction},
    pdfauthor={Benjamin McDonough, Andrea Mari, Nathan Shammah, Nathaniel T. Stemen, Misty, Wahl, William J. Zeng, Peter P. Orth}
}
\usepackage{subcaption}
\usepackage{booktabs}
\usepackage[frozencache=true,cachedir=_minted-Draft-v1]{minted}
\usepackage{svg}
\DeclareMathOperator{\Tr}{Tr}
\usepackage{physics}
\usepackage{algorithm}
\usepackage{algpseudocode}

\usepackage{tikz}
\usetikzlibrary{quantikz}

\definecolor{go_green}{rgb}{0.13, 0.55, 0.13}

\newcommand{\code}[1]{\mintinline{python}{#1}}

\begin{document}
\title{Automated quantum error mitigation based on probabilistic error reduction
\thanks{
This work was primarily supported by the U.S. Department of Energy, Office of Science, National Quantum Information Science Research Centers, Superconducting Quantum Materials and Systems Center (SQMS) under the contract No. DE-AC02-07CH11359 (B.M., P.P.O.).
A.M., N.S., N.T.S., M.W., W.J.Z. were supported by the U.S. Department of Energy, Office of Science, Office of Advanced Scientific Computing Research, Accelerated Research in Quantum Computing under Award Number de-sc0020266 and by IBM under Sponsored Research Agreement No. W1975810.
}
}

\author{\IEEEauthorblockN{Benjamin McDonough}
\IEEEauthorblockA{\textit{Department of Physics and Astronomy} \\
\textit{Yale University}\\
New Haven, Connecticut 06511, USA \\
ben.mcdonough@yale.edu}
\and
\IEEEauthorblockN{Andrea Mari}
\IEEEauthorblockA{\textit{Unitary Fund} \\
San Francisco, California 94104, USA \\
andrea@unitary.fund}
\and
\IEEEauthorblockN{Nathan Shammah}
\IEEEauthorblockA{\textit{Unitary Fund} \\
San Francisco, California 94104, USA \\
nathan@unitary.fund}

\and
\IEEEauthorblockN{Nathaniel T. Stemen}
\IEEEauthorblockA{\textit{Unitary Fund} \\
San Francisco, California 94104, USA \\
nate@unitary.fund}
\and
\IEEEauthorblockN{Misty Wahl}
\IEEEauthorblockA{\textit{Unitary Fund} \\
San Francisco, California 94104, USA \\
misty@unitary.fund}
\and
\IEEEauthorblockN{William J. Zeng}
\IEEEauthorblockA{
\textit{Unitary Fund} \\
San Francisco, California 94104, USA}
\IEEEauthorblockA{
\textit{Goldman \& Sachs} \\
New York, NY, USA\\
will@unitary.fund}
\and
\IEEEauthorblockN{Peter P. Orth}
\IEEEauthorblockA{\textit{Ames National Laboratory} \\
Ames, Iowa 50011, USA\\
}
\IEEEauthorblockA{\textit{Department of Physics and Astronomy} \\
\textit{Iowa State University}\\
Ames, Iowa 50011, USA\\
porth@iastate.edu }
}

\maketitle
\begin{abstract}
Current quantum computers suffer from a level of noise that prohibits extracting useful results directly from longer computations. The figure of merit in many near-term quantum algorithms is an expectation value measured at the end of the computation, which experiences a bias in the presence of hardware noise. A systematic way to remove such bias is probabilistic error cancellation (PEC). PEC requires a full characterization of the noise and introduces a sampling overhead that increases exponentially with circuit depth, prohibiting high-depth circuits at realistic noise levels. 
Probabilistic error reduction (PER) is a related quantum error mitigation method that systematically reduces the sampling overhead at the cost of reintroducing bias. In combination with zero-noise extrapolation, PER can yield expectation values with an accuracy comparable to PEC.
Noise reduction through PER is broadly applicable to near-term algorithms, and the automated implementation of PER is thus desirable for facilitating its widespread use. To this end, we present an automated quantum error mitigation software framework that includes noise tomography and application of PER to user-specified circuits. We provide a multi-platform Python package that implements a recently developed Pauli noise tomography (PNT) technique for learning a sparse Pauli noise model and exploits a Pauli noise scaling method to carry out PER.
We also provide software tools that leverage a previously developed toolchain, employing PyGSTi for gate set tomography and providing a functionality to use the software Mitiq for PER and zero-noise extrapolation to obtain error-mitigated expectation values on a user-defined circuit.
\end{abstract}



\section{Introduction and overview}
\label{sec:intro}
Hardware noise introduces unwanted bias into an expectation value measured on a quantum computer, restricting the applicability of many quantum algorithms on current quantum devices. Combating this problem has led to the development of probabilistic error cancellation (PEC) \cite{temme2017error,endo2018practical, caiQuantumErrorMitigation2022}, a systematic method to remove the noise-induced bias. PEC requires a representation of the desired unitary channel as a linear combination of noisy channels that can be implemented on the hardware, which demands a precise characterization of the noise. Using the linearity of the expectation value, one can then express any ideal value without bias in terms of values obtained from instances of circuits with noisy gates. However, the number of noisy circuits required to represent an ideal circuit increases exponentially with the circuit depth, which would result in an exponentially large number of expectation values being measured. To overcome this issue, the linear combination can be converted into a quasi-probability distribution (QPD), from which sampling circuits yields an unbiased estimator of the value~\cite{temme2017error,endo2018practical, caiQuantumErrorMitigation2022}. 
Due to the presence of negative coefficients in the expansion, the method experiences a sign problem, and requires an exponential sampling overhead to reduce the variance below a desired threshold. 
At current hardware noise levels, the sampling overhead limits PEC to circuits of modest depth~\cite{Zhang2020_NatureComms}. To overcome this restriction, Mari \emph{et al.} developed probabilistic error reduction (PER)~\cite{mari2021extending}, wherein the noise is only partially mitigated rather than fully canceled. By combining partial noise mitigation with virtual zero-noise extrapolation (vZNE), it was suggested that noiseless observables can be approximated at an accuracy similar to that of PEC, but with significantly reduced sampling costs (a related technique was also proposed in Ref.~\cite{ferracinEfficientlyImprovingPerformance2022}). 
\begin{figure}[t]
    \centering
    \includegraphics[width=\linewidth]{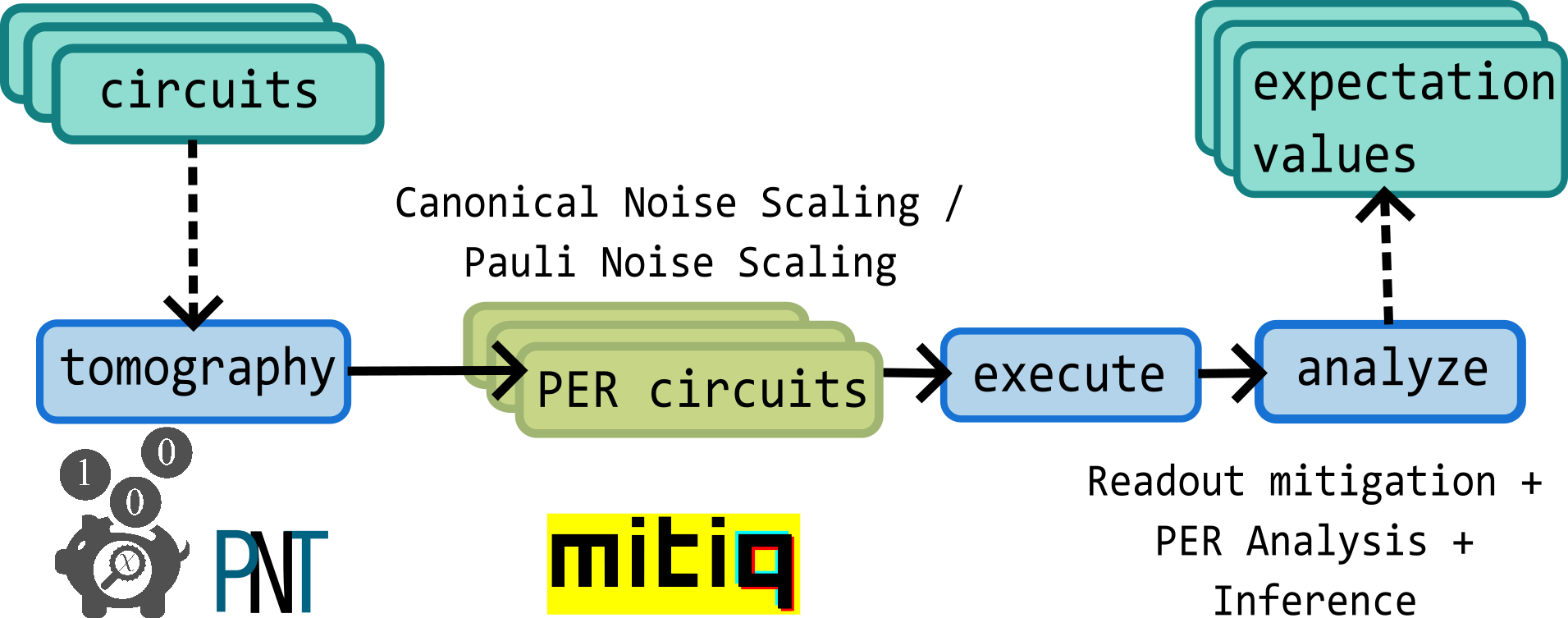}
    \caption{Illustration of automated error mitigation protocol starting from user defined circuits and returning noise-mitigated expectation values. It includes a noise tomography step involving PNT or GST, whose results are used to generate sampled PER circuits via canonical or Pauli noise scaling. The circuits are executed with a user-specified interface to a quantum backend, and the resulting expectation values are adjusted according to the sampling overhead and the PER quasiprobability coefficient of the sampled term. 
    Virtual ZNE is applied to approximate error-mitigated expectation values corresponding to noiseless input circuits.}
    \label{fig:schematic}
\end{figure}

Here, we describe and implement a framework for executing automated quantum error mitigation based on PER. We interconnect hardware noise characterization with the generation, sampling and analysis of PER mitigation circuits to produce expectation values with reduced bias, as illustrated in Fig.~\ref{fig:schematic}. We employ two separate methods of noise characterization: gate set tomography (GST)~\cite{greenbaum2015introduction,Nielsen_2021_Quantum} and a recently proposed cycle benchmarking technique for extracting a sparse Pauli noise model~\cite{flammia2020efficient, berg2022probabilistic}, which we refer to as Pauli noise tomography (PNT). 
GST is implemented using the software package PyGSTi~\cite{nielsen2020probing}, which returns a set of noisy Pauli transfer matrices describing the set of implementable gates $\{\mathcal{O}_\alpha\}$. We convert these to superoperators and pass them into the open-source software package Mitiq~\cite{LaRose2022mitiqsoftware}, a software package containing a collection of quantum error mitigation techniques, to determine the decomposition of a set of desired noiseless unitary superoperators $\mathcal{G}_i$ as
\begin{equation}
    \mathcal{G}_i = \sum_{\alpha} \eta_{\alpha,i} \mathcal{O}_\alpha \,.
    \label{eq:quasi-prob_decomposition}
\end{equation}
Using the previously developed technique of canonical noise scaling~\cite{mari2021extending}, this decomposition is then used to create noise-scaled representations of the channel, $\mathcal{G}^{(\xi)}_i$, that depend on a tunable noise level parameter $\xi$. Circuits are sampled from these representations to obtain several noise-mitigated expectation values that are evaluated at different values of $\xi$. From these results we extrapolate the noiseless ($\xi \rightarrow 0$) expectation value using vZNE. 
Our contribution to this method is to develop software routines to facilitate this workflow of applying PER and vZNE using GST as implemented in PyGSTi for noise tomography to a user-specified circuit. 
We demonstrate this technique in an executable Python notebook, using Mitiq to produce a QPD representation with the noisy operators obtained from GST~\cite{benjamin_mcdonough_2022_7197234}. 

The second method we implement is a recently proposed scheme for efficient benchmarking of a Pauli-twirled noise channel and a technique for sampling from the noise inverse, which we extend to PER.~\cite{berg2022probabilistic,flammia2020efficient}.
This method allows for robust characterization of hardware noise in large devices by keeping the number of measurements constant in the number of qubits. This is possible by placing weak assumptions on the level of correlation in the noise. This method compares favorably with full gate set tomography, which scales exponentially in the number of qubits. 
The PNT method has been used previously to perform PEC on IBM hardware~\cite{berg2022probabilistic}. 
Our key contributions are the following: 1) We combine the advantages of PNT with the sampling overhead reduction of PER and vZNE by extending this method to noise scaling, and 2) We provide a Python package implementing the combined PNT and Pauli noise scaling method as part of the general automated PER framework.
The software automates all steps in the framework: Parsing an arbitrary circuit, generating tomography circuits, collecting and analyzing tomography data to obtain a noise model, generating noise-scaled circuit representations for PER, and running extrapolation to obtain noise-mitigated expectation values. The code is available in the accompanying repository~\cite{benjamin_mcdonough_2022_7197234}. 

In the remainder of the article, we first describe how to combine GST with canonical noise scaling and vZNE and show that this can reduce the sampling overhead compared to PEC. Then, we describe PNT, together with a demonstration of its usage in combination with PER and vZNE. We discuss in detail the developed software package for automated quantum error mitigation based on PNT and Pauli noise scaling to apply PER combined with vZNE for mitigation. Finally, we apply it to mitigate noise in a Trotter dynamics simulation of the transverse field Ising model.
\begin{table}[b]
    \centering
    \begin{tabular}{*{2}{c}} 
     \toprule
     Acronym & Meaning \\
     \midrule
     PEC & Probabilistic error cancellation \\ 
     PER & Probabilistic error reduction \\ 
     GST & Gate set tomography \\
     PNT & Pauli noise tomography \\
     vZNE & virtual Zero-Noise Extrapolation \\
     QPD & Quasi-probability distribution \\
     \bottomrule
    \end{tabular}
    \caption{List of acronyms used in the text.}
    \label{tab:acronyms}
\end{table}

\section{Gate set tomography and and probabilistic error reduction} 
\subsection{Gate set tomography}
Gate set tomography (GST)~\cite{greenbaum2015introduction,Nielsen_2021_Quantum} is a method for characterizing noise associated with a set of gate operations capable of preparing a complete set of density matrices. The details of the long-sequence GST used here, including the choice of gate strings that maximally amplify the errors and a choice of suitable SPAM gates, are implemented by the software package PyGSTi~\cite{nielsen2020probing}. PyGSTi provides model packs with pre-computed gate strings and SPAM gates. Here, we use the \textit{sm1Q-XZ} model pack to reconstruct the single qubit gate set $\mathcal G = \{ RX(\frac{\pi}{2}), RZ(\frac{\pi}{2}) \}$ on Rigetti Aspen-11, a cloud-accessible superconducting-circuit quantum processing unit (QPU). The maximum depths for the long sequence gate strings were chosen to be $\{2,4,8,16,32\}$, and the GST experiment include a total number of $550$ circuits, each of which was run at $1000$ shots. 

The output of PyGSTi is the set of operators represented as Pauli-transfer matrices (PTMs), which can be further processed as part of an error mitigation workflow using Mitiq. GST results exhibit a gauge degree of freedom related to the uncertainty in both state-preparation and measurement processes, which results in a set of PTMs which are similar to the ideal operations up to conjugation by an invertible matrix.
The gauge freedom is by definition not detected in the measurement of expectation values, but it does affect the decomposition of ideal gates in terms of noisy ones in Eq.~\eqref{eq:quasi-prob_decomposition} and thus the sampling overhead of PEC and PER. 
Here, we use the knowledge that the $RZ$ gate is virtualized on many platforms, including the Rigetti hardware, and use PyGSTi's inbuilt gauge optimization methods to choose a gauge such that the $RZ$ PTM is noiseless.
Full gate set tomography scales exponentially in complexity with the number of qubits, which in practice restricts the number of qubits operated on by a gate set to below three. An avenue for future work is to 
explore variants of GST protocols capable of characterizing noise in larger devices~\cite{Song_2019_Science_Advances,Nielsen_2021_Quantum}.
This is typically made possible by placing stronger assumptions on the locality of the noise, which may limit the effectiveness of these protocols for PEC due to the untracked level of crosstalk in the device~\cite{harperEfficientLearningQuantum2020,mckayCorrelatedRandomizedBenchmarking2020}. 

\subsection{Canonical noise scaling and vZNE}
\label{subsec:PER_canonical_noise_scaling}
\begin{figure}[t]
    \centering
    \includegraphics[width = \linewidth]{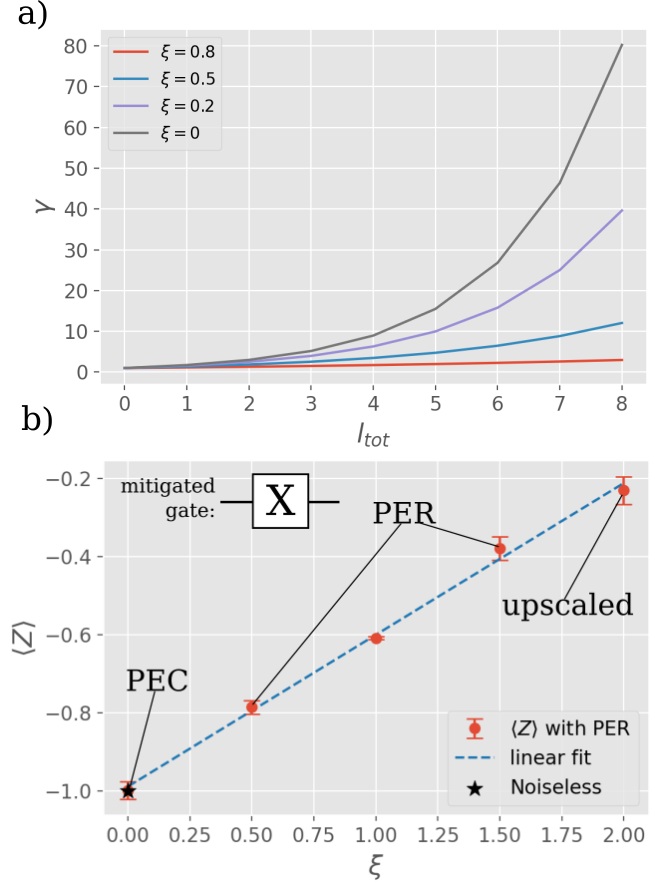}
    \caption{(a) Sampling overhead $\gamma^{(\xi)}_{\text{tot}} = (\gamma^{(\xi)})^l$ versus circuit depth $l$ for different noise strengths $\xi$ using a canonical noise scaled representation derived from GST. Here, $\gamma_0 = 1.73$ as obtained from GST on Rigetti Aspen-11 and $\gamma^{(\xi)}$ is defined in Eq.~\eqref{eq:gamma_PER}. The figure demonstrates that a realistic sampling overhead $\gamma$ of PEC on current hardware can be prohibitive for circuits of even modest depth $l_{tot}$, which validates the need for methods to reduce the overhead of noise-sensitive error mitigation such as PER and vZNE.
    (b) PER results for $\langle Z \rangle$ after application of a noisy $X$ gate to $\ket{0}$. This was simulated for a random Pauli noise model chosen to have an overhead matching that obtained from GST on the Rigetti QPU. The decomposition of the noise reduced operator $\mathcal{G}^{(\xi)}$ in Eq.~\eqref{eq:PER_decomposition} can be obtained from GST using Mitiq's optimal representation algorithm and applying canonical noise scaling or from partially inverting the Pauli noise model $\Lambda$ obtained from PNT (discussed in detail in Sec.~\ref{subsec:PNT}). 
    }
    \label{fig:overhead_and_zne}
\end{figure}

Noise characterization is only the first step for noise-sensitive quantum error mitigation. To proceed, we first convert the PTMs of the noisy gate operations obtained from GST into a superoperator representation.  
We then feed these into Mitiq to obtain a quasi-probability representation of ideal noiseless gates, as shown in Eq.~\eqref{eq:quasi-prob_decomposition}. Applying canonical noise scaling~\cite{mari2021extending} to this representation, it is then straightforward to derive a decomposition of noise-reduced gates $\mathcal{G}^{(\xi)}_i$ with noise strength $\xi \in [0, \frac{\gamma_i+1}{\gamma_i-1}]$:
\begin{equation}
    \mathcal{G}^{(\xi)} = (\gamma^+ - \xi \gamma^-) \Phi^+ - (1 - \xi) \gamma^- \Phi^- \,.
    \label{eq:PER_decomposition}
\end{equation}
Here we have dropped the subscript $i$ and separated the positive and negative coefficients $\eta_{\alpha}$ in Eq.~\eqref{eq:quasi-prob_decomposition} into $\gamma^+ = \sum_{\eta_\alpha>0}|\eta_{\alpha}| > 0$ and $\gamma^- = \sum_{\eta_\alpha < 0} |\eta_\alpha|$, corresponding to the positive and negative volumes of the QPD. We defined $\gamma = \gamma^+ + \gamma^- = 1 + 2 \gamma^-$, which is the sampling overhead at $\xi = 0$, corresponding to PEC. The overhead is determined by $\gamma^-$, which is referred to as the negativity of the QPD. 
We also introduced the completely-positive and trace preserving (CPTP) maps $\Phi^{\pm} = \sum_{\eta_\alpha \gtrless 0} \frac{|\eta_\alpha|}{\gamma^\pm}\mathcal{O}_\alpha$. The sampling overhead for $\xi \in [0,1]$ is given by 
\begin{equation}
    \gamma^{(\xi)} = \gamma - \xi (\gamma - 1) \,.
    \label{eq:gamma_PER}
\end{equation}
This shows that PER provides a way to systematically reduce this overhead by removing the noise only partially. This reduction in overhead is shown explicitly in Fig.~\ref{fig:overhead_and_zne}(a) for a noise-scaled reconstruction of an $RX(\frac{\pi}{2})$ gate with a noisy single-qubit gate set obtained using GST on Rigetti's Aspen-11, which yields $\gamma = 1.73$. The sampling overhead for a circuit of modest depth of $l=8$ gates, $(\gamma^{(\xi)})^l$, is reduced from over 80 for PEC $(\xi=0)$ to about $5$ for $\xi = 0.8$. We note that enforcing a noiseless $RZ$ gate increases the overhead to $\gamma = 2.67$, leading to an even more steeply increasing sampling cost. 

As shown in Fig.~\ref{fig:overhead_and_zne}(b), although the price of reducing the overhead is additional bias, we can improve the estimate of the desired noiseless expectation values by leveraging the fact that expectation values converge to their ideal values at $\xi = 0$. Several values obtained through PER at different noise levels $\xi$ can thus be used to extrapolate to the zero-noise limit. This vZNE procedure~\cite{mari2021extending} can yield a zero noise estimator close to PEC at much lower sampling costs. 

Finally, we note that Eq.~\eqref{eq:PER_decomposition} can also be used to scale up the noise when choosing $\xi \in [1, \frac{\gamma_i+1}{\gamma_i-1}]$, which does not incur any additional sampling overhead. Within ZNE one can thus combine estimator values at both reduced and increased noise levels, which can further improve the ZNE fit at low additional sampling costs (see Fig.~\ref{fig:overhead_and_zne}).

\section{Pauli noise tomography and probabilistic error reduction}
\label{sec:PNT_PER}
We provide software that implements a recent noise characterization protocol based on cycle benchmarking~\cite{flammia2020efficient,berg2022probabilistic}, which we refer to as Pauli noise tomography (PNT). This technique involves applying a Pauli twirl surrounding the Clifford entangling gates contained in layers of the circuit. This has been shown to convert an arbitrary noise channel into a Pauli channel\cite{wallman2016noise,flammia2020efficient, berg2022probabilistic}. The resulting Pauli fidelities characterize the twirled noise. PNT is able to achieve a constant scaling in the number of qubits, which makes this benchmarking method practically applicable to larger systems. This is achieved in part by modeling the twirled noise channel with a Lindbladian whose quantum jump operators are proportional to Pauli terms with support on qubits that are physically connected to each other in the quantum processing unit (QPU). This physically motivated assumption greatly reduces the complexity of the algorithm. After extracting a sparse Pauli noise model, PEC circuits can be generated using an efficient procedure to sample from the inverse of the noise channel~\cite{berg2022probabilistic}. We here extend this approach to PER by constructing a \emph{partially inverted} noise channel, enabling the level of noise to be controlled by a parameter $\xi$. We demonstrate the efficacy of this method in an accompanying tutorial notebook, where it is applied to a Trotter dynamics simulation of four qubits on a noisy backend emulator~\cite{benjamin_mcdonough_2022_7197234}. 

\subsection{Pauli noise tomography}
\label{subsec:PNT}
The advantage of PNT is its ability to capture correlated noise with low algorithmic complexity. The method works 
by decomposing circuits into ``dressed" layers, which consist of a sequence of layers of single-qubit gates that can be compiled together, followed by a layer of self-adjoint Clifford gates with disjoint support. On platforms that support single-qubit rotations and the CNOT or CZ gate, any circuit can be decomposed into this form.

This reduction in complexity is achieved by first Pauli-twirling the 
noise channel $\Lambda_l$ associated with a layer $l$. The twirl is implemented by sampling random Pauli operators 
around the noisy layer such that they have no logical effect on the circuit. This has the effect of diagonalizing $\Lambda$ (dropping the subscript $l$) in the Pauli basis. This can be expressed symbolically as
\begin{equation}
    \mathbb{E}_i[P_i\Lambda(P_i\rho P_i) P_i] = \sum_{k} c_k P_k \rho P_k \equiv \Lambda^{\mathbb{P}^n}(\rho)\,.
\end{equation}
Once the channel is diagonalized, it can be characterized by the fidelities $f_a$, which are the diagonal elements of the Pauli transfer matrix:
\begin{equation}
    f_a = \frac{1}{d}\Tr(P_a\Lambda^{\mathbb{P}^n}(P_a)) \,.
\end{equation}
For a Pauli channel, these fidelities can be easily measured. If $\ket{+}_a$ is a $+1$ eigenstate of $P_a$, then we observe
\begin{equation}
    \frac{1}{d}\Tr(P_a\Lambda^{\mathbb{P}^n}(P_a)) = \Tr(P_a\Lambda^{\mathbb{P}^n}(\ketbra{+}{+}_a)) \,.
\end{equation}
Since the Pauli matrices $P_a$ are eigenvectors of the twirled noise channel (dropping the $\mathbb{P}^n$ superscript from here on), the fidelities can be determined through exponential fits. However, since the noise is attached to a self-adjoint layer of Clifford gates $\mathcal{C}$, applying an even number of these layers results in the measurement of fidelity pairs:
\begin{equation}
    \frac{1}{d}\Tr(P_a(\Lambda\circ \mathcal C)^{2n}(\rho)) = (f_a f_{a}')^n\,.
\end{equation}
where $f_{a}'$ is the fidelity of $P_a^C = \mathcal{C}(P_a) = C P_a C^\dag$. In the most general case, there are still exponentially many fidelities $f_a$ requiring measurement. Under the assumption that noise correlations are strongest when there exist physical connections between qubits, the twirled noise is modeled using a master equation where the quantum jump operators $\sqrt{\lambda_k} P_k$ are chosen to be only those Paulis which have support on neighboring qubits~\cite{berg2022probabilistic}. It can be shown that the noise channel, written as a superoperator, assumes the following form:
\begin{equation}
    \Lambda = \prod_k (w_k \mathcal I + (1-w_k)\mathcal P_k) \,,
    \label{eq:Pauli_noise_model}
\end{equation}
where $w_k = \frac{1}{2}(1+e^{-2\lambda_k})$, $\mathcal{I}$ is the identity, and $\mathcal{P}_k \rho = P_k \rho P_k^\dag$. Given this form of the noise, the fidelitites $f_a$ can be related to the coefficients $\lambda_k$ by the equation
\begin{equation}
f_a = \prod_{\{P_a, P_k\}=0} (1-2w_k) = \exp(-2\sum_{\{P_a, P_k\}=0}\lambda_k) \,.
\label{eq:coefficients_and_fidelities}
\end{equation}
Only the fidelities $f_a$ corresponding to the terms in the sparse model require measurement to reconstruct the model. Many QPU architectures exhibit a local qubit connectivity, often restricted to nearest-neighbor pairs for which the number of terms in the model scales linearly in the number of qubits. The ability to simultaneously measure commuting terms results in the additional improvement to constant scaling. Since PNT characterizes the noise that is associated with a layer of Clifford gates, methods such as Trotter time evolution or variational quantum circuits that involve repetition of identical circuit layers also exhibit a constant scaling in the depth of the circuit. 

\begin{figure}     
\centering
\includegraphics[width=\linewidth]{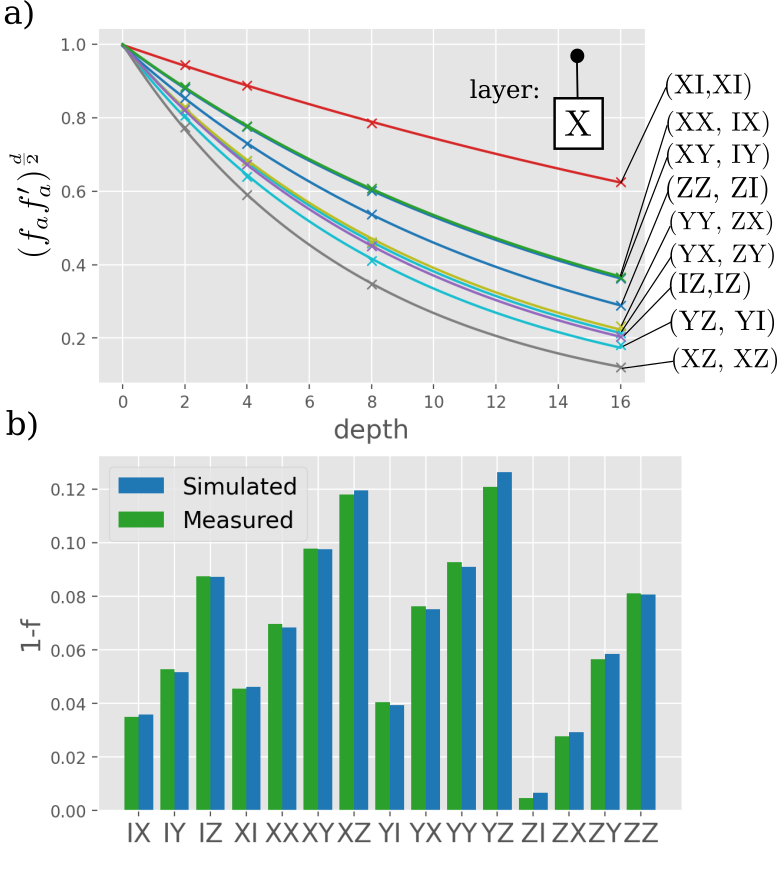}
    \caption{PNT results for a layer consisting of a CNOT gate under a randomly generated Pauli noise model composed with an amplitude damping noise channel with $p=0.01$.
    We use $32$ samples for the Pauli twirl and run different circuits depths for $250$ shots each. Panel (a) shows that the fidelity pairs $f_a f_a'$ of all Pauli operators on two qubits decay exponentially with circuit depth. The amplitude damping noise was included to show that the twirling properly diagonalizes the channel as evidenced by the exponential decays observed in the figure. 
    Panel (b) compares the measured fidelity pairs to the products of diagonal elements of the twirled channel transfer matrix. The agreement between the measured and ideal values shows the efficacy of PNT.}
    \label{fig:CNOT_layer}
\end{figure}

In Fig.~\ref{fig:CNOT_layer} we show PNT results for a randomly generated Pauli noise model. Since Pauli twirling is shown to convert arbitrary noise into Pauli noise, this efficacy extends to more realistic noise models as well. We observe excellent agreement between the simulated fidelities and the fidelities obtained via PNT for a layer consisting of a CNOT gate. 
Once the fidelities $f_a$ are obtained, a matrix $M$ can be constructed via $[M]_{ab} = \langle P_a, P_b\rangle_{sp}$, where $\langle \cdot, \cdot \rangle_{sp}$ refers to the symplectic inner product, which is zero if $[P_a,P_b] =0$ and one otherwise. The vector of fidelities $\boldsymbol{f}$ is then related to the vector of noise model parameters $\boldsymbol{\lambda}$ via 
\begin{equation}
2 M \boldsymbol{\lambda} + \ln(\boldsymbol{f}) = 0 \,.
\label{eq:solve_noise_model}
\end{equation}
The $\lambda_k$ can be approximated from the measurements of $f_a$ using a non-negative least squares algorithm, yielding a sparse Pauli noise model parametrized by $\lambda_k$ (or equivalently $w_k$).

\subsection{Partial Pauli noise inversion}
\label{subsec:PER_partial_noise_inversion}
We incorporate PNT into the PER error mitigation framework by generalizing the noise inversion through sampling from a QPD representation as discussed in \cite{berg2022probabilistic} to performing a partial inverse. A partial inverse of the noise channel can be constructed with the form
\begin{equation}
    \Lambda^{(\xi)} = \gamma^{(\xi)}\prod_k \Bigl( w_k^{(\xi)}\mathcal I +\operatorname{sgn}(\xi-1) (1-w_k^{(\xi)})\mathcal P_k\Bigr)\,.
    \label{eq:Lambda_lambda}
\end{equation}
Here, $w^{(\xi)} \equiv \frac{1}{2}(1+e^{-2|1-\xi|\lambda_k})$ and the sampling overhead 
\begin{equation}
\gamma^{(\xi)} =
\begin{cases}
\exp[2(1-\xi)\sum_{k}\lambda_k] & \xi < 1\\
1 & \xi \geq 1
\end{cases}
\label{eq:overhead}
\end{equation}
quantifies the variance in the estimator. This channel $\Lambda^{(\xi)}$ exhibits similar useful properties as we observed within canonical noise scaling in Sec.~\ref{subsec:PER_canonical_noise_scaling}. For example, upscaling the noise does not incur any sampling overhead yet can still provide useful information for vZNE.
For intermediate values of the noise $0 < \xi < 1$, the overhead interpolates exponentially between unity and the PEC value, which enables a substantial reduction in the number of circuits that need to be executed. When $\xi \to 0$, we have $\Lambda^{(\xi)} \to \Lambda^{-1}$, corresponding to PEC. On the other hand, when $\xi \to \infty$, $\Lambda^{(\xi)}$ approaches maximal depolarizing noise, which maps every density matrix to the identity. Lastly, the product form of this partial inverse ensures that this method retains the efficient sampling from the inverse discussed in \cite{berg2022probabilistic}. The sampling procedure is described in Algorithm~\ref{alg:PER}.

\begin{algorithm}
    \caption{Description of PER routine}\label{alg:PER}
    \textbf{Input:} Circuit with layers $l \in \{1, .., l_{tot}\}$, each with noise model parameters  $\{w_{l1}^{(\xi)}, ..., w^{(\xi)}_{ln}\}$

    \textbf{Output:} A sample of the PER expectation value (before readout error mitigation)
    \begin{algorithmic}[1]
    \State Let $\alpha \equiv 1$
        \For{$l \in \{1, \ldots, l_\text{tot}\}$}
            \State Compose layer $l$ into circuit
            \For{$k \in \{1,\ldots, n\}$}
                \State Sample $I$ with probability $w_{lk}^{(\xi)}$ and $P_{k}$ otherwise
                \State Multiply $\alpha$ by $\gamma_l^{(\xi)}$
                \If{$P_k$ was sampled}
                    \State Multiply $\alpha$ by $-1$
                \EndIf
            \State Compose sampled operator into circuit
        \EndFor
    \EndFor
        \State Run the circuit and get the expectation value
        \If{$\xi < 1$}
            \State Scale result by $\alpha$
        \EndIf
    \end{algorithmic}
\end{algorithm}
The total overhead is a product of the overhead of the individual layers $l$, $\gamma_{\text{tot}}^{(\xi)} = \prod_{l} \gamma_l^{(\xi)}$, and so it still scales exponentially in the number of layers. However, by making this exponential scaling weaker, one can extend the practical application of error mitigation to larger circuits. Implementation details are discussed in the following section.

\section{Software for automated error mitigation} 
\label{sec:software_tools}
\begin{figure*}
    \centering
    \includegraphics[width = \linewidth]{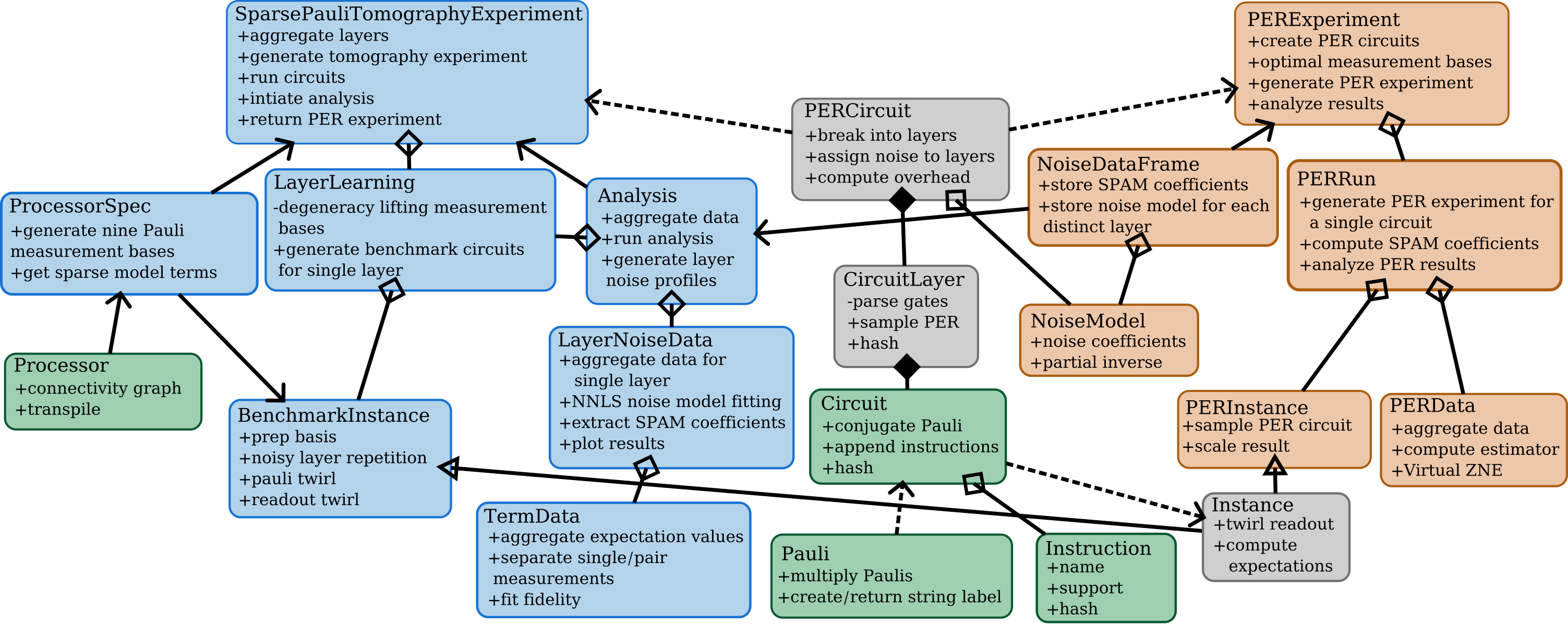}
    \caption{UML diagram describing software tools for automated error mitigation. The solid lines with arrowheads represent association, the dotted lines with solid arrowheads represent dependency, the solid lines with triangular heads represent implementation, the lines with closed diamond heads represent composition, and the open diamond heads represent aggregation. Objects involved in tomography are shown in blue, objects in orange are involved in PER, and gray objects are used by both. The green objects are wrappers for external implementations (such as Qiskit) and require implementations with external dependencies.}
    \label{fig:uml_diagram}
\end{figure*}
To make the techniques described above practical for the end user, we here  
present a software package to automate the implementation of PER, from tomography to extrapolation. Our first attempt at such an automated framework is an object-oriented interface written in Python automating the PNT and Pauli noise scaling process. 
The chief goal of this implementation is to allow the user to apply this technique in its entirety without being burdened by the details. This section describes the functionality of this package to automate the process of performing tomography and using it to carry out PER, ultimately obtaining an error-mitigated set of desired expectation values. 

At the top level, the software is divided into two parts: (i) tomography and (ii) PER.  The program is intended to be easily extended to different platforms such as PyQuil or Circ, interacting with the native implementation through the abstract classes \mintinline{python}{Circuit}, \mintinline{python}{Processor}, \mintinline{python}{Pauli}, and \mintinline{python}{Instruction}. A Unified Modeling Language (UML) diagram highlights these objects (green) in Fig.~\ref{fig:uml_diagram}. 
 These classes are wrappers for objects and behaviors common to many quantum development toolkits, and can be overridden to provide support for another API's. This allows circuits to be run in their native representation without any conversion. Currently only the Qiskit interface has been implemented. 

\subsection{Pauli Noise Tomography software tools}
\label{susec:software_PNT}
The method used for PNT is described in Ref.~\cite{berg2022probabilistic}. To begin the process, the \mintinline{python}{SparsePauliTomographyExperiment} class is initialized with a list of circuits, a mapping of algorithm qubits to physical qubits, and a quantum backend. The experiment class initializes an instance of \mintinline{python}{ProcessorSpec}, which uses the coupling map from the backend to generate a list of the Pauli terms with support on neighboring qubits. Using the sweeping algorithm described in Ref.\cite{berg2022probabilistic}, this object chooses the nine optimal measurement bases from which simultaneous measurements can be used to obtain fidelities of all Pauli terms in the sparse model. Then, the \mintinline{python}{PERCircuit} class separates each circuit into layers that have the form of any number of single-qubit gates followed by a layer of self-adjoint Clifford gates with disjoint supports.
The noise of one of these layers is assumed to be determined just by the Clifford gates in the layer. The self-adjoint Clifford layers, from here on just ``Clifford layers," are therefore hashed into a set to be benchmarked individually.

Next, the user can call the \mintinline{python}{generate} method on the experiment class. The generation procedure creates a \mintinline{python}{LayerLearning} object for each distinct Clifford layer, which are responsible for generating the benchmark circuits for tomography. Each learning procedure consists of two types of circuits. The first represents the measurement of fidelity pairs. These are the measurements that can be made at multiple depths and fit to exponential decays. When the fidelities of two model terms appear in a pair, a degeneracy is created in the model. This degeneracy is lifted by the second type of circuit, which consists of a single repetition of the noisy layer. The use of a single repetition causes the preparation and measurement bases to differ, which makes this type of measurement less resistant to state preparation errors. 
\begin{figure}[t!]
    \centering
    \includegraphics[width=.75\linewidth]{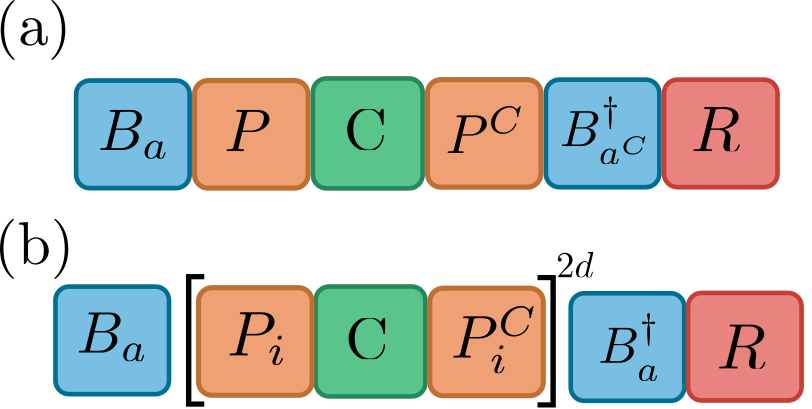}
    \caption{Illustration of the two forms of benchmark circuits. The $B_a$ gates change from the computational basis into the Pauli basis $P_a$ being benchmarked. The Pauli twirl operator $P$ is sampled at random from the Pauli group. $C$ is the Clifford layer, and the superscript $(\cdot)^C$ represents conjugation by the Clifford layer. The $R$ gate is readout twirling, sampled at random from $\{I, X\}^{\otimes n}$ \cite{Karalekas_2020, berg2020model}. Adjacent single-qubit gates are compiled together in a way that preserves the structure of the dressed layers. Panel (a) shows single-depth measurements that lift the degeneracy in the model. Panel (b) shows a noisy layer repeated an even number of times, resulting in benchmarking fidelity pairs.}
    \label{fig:benchmark_circuits}
\end{figure}
In addition to the nine Pauli bases chosen for the pair measurements, there are at most six bases that are needed to make the degeneracy-lifting measurements, independent of the number of qubits. The initialized \mintinline{python}{LayerLearning} objects choose these basis in relation to the corresponding Clifford layer. 

Each \code{LayerLearning} generates a list of \mintinline{python}{BenchmarkInstance} objects representing the desired measurements, and each \mintinline{python}{BenchmarkInstance} generates the circuit corresponding to this instance. This includes readout twirling, following the method described in \cite{berg2020model}, where gates are randomly sampled from the set $\{I,X\}$ before measurement on each of the qubits, and then the twirl is inverted in software. This has the effect of diagonalizing the readout error in the computational basis. The basis preparation, measurement, and readout twirling gates are stored as metadata.
The next step for the user is to call the \mintinline{python}{run} method on the experiment. This method accepts as input a user-defined ``executor" function, which executes a list of circuits on the QPU and returns the results as a dictionary.

To complete the tomography procedure, the \code{analyze} method can be called.  This initializes the \code{Analysis} class, which is constructed with the \code{LayerLearning} classes from the experiment containing the experiment parameters and the benchmark data. The \code{Analysis} class creates a dictionary of \code{LayerNoiseData} objects for processing the data associated with each distinct Clifford layer. Each of these has a list of \code{TermData} objects, which will store the expected data for each term in the sparse model. 
The \code{get_expectation(pauli)} method is called on each \code{BenchmarkInstance} and used to update the estimator in the \code{PauliTerm} objects for each of the measurements that can be made simultaneously. The \code{get_expectation(pauli)} method is responsible for untwirling the result of the readout using the stored metadata and returning the overlap with the desired Pauli term in the computational basis. Finally, the resulting values are sorted in each \code{TermData} object into the type and depth of the measurement, where ``type" indicates whether the measurement is a fidelity pair or a single-depth measurement.

Once the \code{TermData} objects for each \code{LayerNoiseData} have been populated, the results can be used to fit the fidelities. The pair fidelities are determined first through fits to $ae^{-b}$, where $a$ quantifies the combined SPAM errors and $e^{-b} = \sqrt{f_af_a'}$ is the square root of the fidelity pair. 
The pair measurements are used to determine SPAM coefficients for each measurement basis in order to mitigate SPAM errors in the single-depth measurements. The degeneracy-lifting fidelities are determined directly from the estimators of the expectation values, and the SPAM coefficients from the fits are used to reduce SPAM errors.
Since the pair measurements are assumed to be more accurate, the value of a single-depth measurement $f_a$ is limited by the pair fidelity $f_a'$ via the constraint $f_af_a' \leq 1$ and an exception is logged if $f_a$ violates this bound. 

After this set of single and pair fidelities have been determined, each layer can be fit to the sparse noise model to determine the layer coefficients. This is done as described in \cite{berg2022probabilistic}: one forms a vector $\boldsymbol{b}$ from the fit results, and forms lists $F_1$ and $F_2$ containing Pauli operators. The entries of $\boldsymbol{b}$ are either single fidelities $b_a = f_a$ or fidelity pairs $b_a = \sqrt{f_a f_a'}$. If $b_a$ is a pair, then $P_a$ is added to $F_1$ and $P_a'$ is added to $F_2$. If $b_a$ is not a pair, then $P_a$ is added to both $F_1$ and $F_2$. From here, one constructs matrices $M_1$ and $M_2$ using the definitions $[M_1]_{ab} = \langle F_{1a}, F_{1b}\rangle_{sp}$ and $[M_2]_{ab} = \langle F_{2a}, F_{1b}\rangle_{sp}$.

The Pauli noise model coefficients $\boldsymbol{\lambda}$ are obtained from a numerical solution of [cf. Eq.~\eqref{eq:solve_noise_model}]
\begin{equation}
    (M_1 + M_2)\boldsymbol{f} + \ln(\boldsymbol{\lambda}) = 0 \,,
\end{equation}
where the logarithm is taken elementwise. The result of the fit is a \code{NoiseModel} object, and all of these are composed in a \code{NoiseDataFrame} object, which stores them as a dictionary with Clifford layers as keys. In addition, the SPAM coefficients for each single-weight Pauli measurement are averaged together and stored to model the readout error for use in mitigation. This \code{NoiseDataFrame} object forms the link between the tomography and PER portions of the protocol.

The software package also provides several visualization tools to plot results of different steps of the process. Within the tomography procedure, there are three different plots implemented. The first is the plot of the exponential decays of the Pauli fidelity pairs with increasing circuit depth. This can be used to see if enough samples were taking from the twirl to properly diagonalize the channel. Next, the infidelities of Pauli operators can be plotted by specifying a list of qubits as single-element tuples or qubit pairs as tuples. Lastly, the coefficients $\lambda_k$ appearing in the generator can be plotted against each other. This can be especially helpful to compare the errors experienced by different qubits and identify the dominant sources of error.

\subsection{Probabilistic error reduction software tools}
\label{subsec:software_PER}
The \code{NoiseDataFrame} object resulting from the tomography contains all the data needed to carry out PER. Calling the \code{create_per_experiment(circuits)} method on the experiment class passes this object to a new PER experiment with a set of desired circuits to mitigate. Upon initialization, the \code{PERExperiment} class passes each circuit to the \code{PERCircuit} class, which breaks each circuit into dressed layers of the form described in Sec.~\ref{subsec:PNT}.

The generation of the PER circuits is initiated by calling \code{generate} on the \code{PERExperiment} instance. The arguments of this method are the desired expectation values, the number of samples to take from the combined distribution of the partial noise inverse, the Pauli twirl, and the readout twirl, and finally, the different noise strengths at which to run the circuits. Then, the minimal set of measurements that can simultaneously reconstruct the desired expectation values is determined. A new \code{PERRun} class is instantiated for each of the circuits on which to run PER. Each \code{PERRun} object is responsible for creating the desired \code{PERInstances} representing the collection of a simultaneous subset of the desired expectation values at a particular noise strength. The sampling procedure is described in Sec.~\ref{subsec:PER_partial_noise_inversion}.
\begin{figure}[t!]
    \centering
    \includegraphics[width=.85\linewidth]{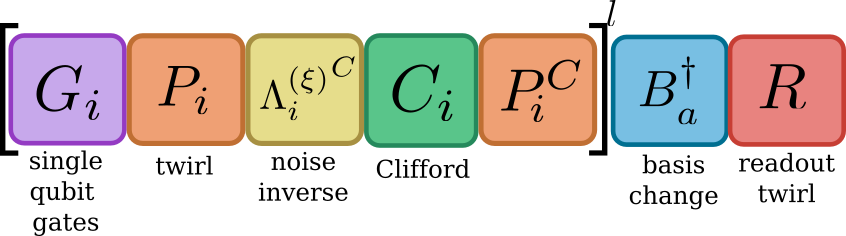}
    \caption{Illustration of PER circuits. The user-specified circuit is parsed to obtain the single-qubit gates $G_i$ and the Clifford layer $C_i$ for each layer $i$ in the circuit. Each layer is assigned a noise model produced by the tomography. $B_a$ represents the gates used to change from the computational basis to the eigenbasis of $P_a$. The inverse is sampled at the desired noise strength along with the Pauli and readout twirling. The symbol $P^C = CPC^\dag$ denotes the conjugation of $P$ by the Clifford layer $C$.}
\end{figure}

Once the circuits have been generated, they can be executed using the same \code{executor} method used in the tomography section. The result of running each circuit is paired with the \code{PERInstance} that produced it for later analysis. Once the run is complete, the \code{analyze} method can be invoked on the PER experiment to process the data. This calls the \code{analyze} method on each of the \code{PERRuns} in the experiment, which in turn assigns the populated \code{PERInstance} objects to a \code{PERData} object for each expectation value that can be simultaneously determined from the instance. 

Object \code{PERData} calls \code{get_adjusted_expectation} on each of the \code{PERInstance} objects. This method is responsible for converting the resulting expectation value into a PER estimator by rescaling the raw expectation value of the circuit with the sign recording the parity of the number of nonidentity operators sampled in Eq.~\eqref{eq:Lambda_lambda} and the overhead $\gamma_l^{(\xi)}$ corresponding to the circuit layer to which it belongs. At this point, the SPAM coefficients obtained from tomography are used to perform readout error mitigation on the expectation value. Readout error mitigation is currently implemented under the assumption of uncorrelated readout noise to cut down on the number of circuits that need to be run, but as work such as \cite{berg2020model} suggest, this assumption may be too strong, and this functionality should be improved in the future.

Finally, each \code{PerData} object performs vZNE on the expectation values taken at different noise strengths to yield a final PER estimate of the desired expectation values on the list of input circuits. The ansatz function used for the fit is $ae^{-b}$, where $a$ is the ideal expectation value. There is reason to believe that this ansatz is at least approximately accurate for any circuit, but this merits future exploration.

The principal plotting tool in the PER module is the ability to show the convergence of the expectation value for different strengths of noise against the exponential fit. The \code{analyze} method returns the \code{PERRun} objects corresponding to different circuits, and the \code{get_result(pauli)} method can be used to obtain the \code{PERData} for a specific expectation value on the desired circuit. This object contains the data and plotting tools for this run.

\section{Application tutorial}
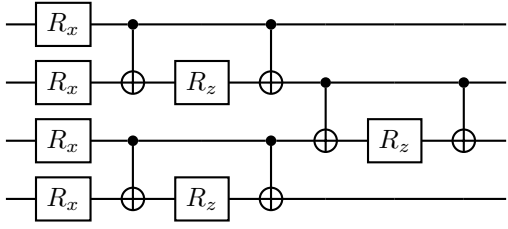
\begin{figure}[t!]
    \begin{center}
        \begin{quantikz}[row sep=0.2cm, column sep=0.4cm]
            & \gate{R_x} & \ctrl{1} & \qw        & \ctrl{1}  & \qw      & \qw        & \qw      & \qw \\
            & \gate{R_x} & \targ{}  & \gate{R_z} & \targ{}   & \ctrl{1} & \qw        & \ctrl{1} & \qw \\
            & \gate{R_x} & \ctrl{1} & \qw        & \ctrl{1}  & \targ{}  & \gate{R_z} & \targ{}  & \qw \\
            & \gate{R_x} & \targ{}  & \gate{R_z} & \targ{}   & \qw      & \qw        & \qw      & \qw
        \end{quantikz}
        \caption{The realization of a single Trotter step as a quantum circuit. Here we have defined $R_x \equiv RX(-2h\delta t)$ and $R_z \equiv RZ(2J\delta t)$}
        \label{fig:trotterstep}
    \end{center}
\end{figure}

We choose as our practical application a Trotter simulation of the postquench dynamics in the one-dimensional transverse-field Ising model (TFIM) with Hamiltonian
\begin{equation}
    H = -J\sum_j Z_jZ_{j+1}-h\sum_j X_j \,.
\label{eq:TFIM}
\end{equation}
This example is used for comparison to Ref.~\cite{berg2022probabilistic}.
A single Trotter step can be constructed as a \code{QuantumCircuit} object in the form of Fig.~\ref{fig:trotterstep}. 
To simulate the evolution at different points in time, a method can be created to repeat this Trotter step $n$ times by defining a function \code{trotterCircuit(n)}.
A list of circuits corresponding to increasing Trotter steps can be generated:
\begin{minted}{python}
circuits = [
    trotterCircuit(n) for n in range(0, d)
]
\end{minted}
Next, a backend should be initialized and used to transpile the circuits. In the tutorial notebook, we use \code{FakeVigoV2}. 
PNT is initialized via
\begin{minted}{python}
from tomography.experiment \
import SparsePauliTomographyExperiment \
as pnt
experiment = pnt(
    circuits=circuits,
    inst_map=[0,1,2,3], 
    backend=backend,
)
\end{minted}
The transpiled circuits are passed to the experiment, along with the backend and a map from the virtual qubits in the circuit to the physical qubits on the backend. Next, the circuits for tomography can be generated with
\begin{minted}{python}
experiment.generate(
    samples=32, 
    single_samples=200, 
    depths=[2,4,8,16],
)
\end{minted}
These are the default options for the parameters. Increasing the samples will increase the goodness of the fit (for details see Sec.~\ref{sec:software_tools}).
The execution of circuits is exposed to the user by the use of a method which takes a list of quantum circuits and returns the result of executing these circuits on a quantum computer as a counts dictionary. An example of such an executor is
\begin{minted}{python}
def executor(circuits):
    job = backend.run(circuits)
    return job.result().get_counts()
\end{minted}
This executor can be used to call the \code{run} method of the experiment:
\code{experiment.run(executor)}
Once the execution is finished, the data can be analyzed by calling
\begin{minted}{python}
experiment.analyze()
\end{minted}
The experiment is now populated with all of the noise data needed to carry out PER. With this, the PER experiment can be set up for the desired circuits by calling
\begin{minted}{python}
perexp = \
experiment.create_per_experiment(circuits)
\end{minted}
The value we want to mitigate is the $z$-component of the magnetization, $M_z = \frac{1}{N} \sum_{i=0}^{N-1} \langle Z_i \rangle$. We can collect these expectation values by passing them to the \code{generate} method:
\begin{minted}{python}
expectations = ["ZIII", "IZII",
                "IIZI","IIIZ"]
perexp.generate(
    expectations, 
    samples=1000, 
    noise_strengths=[0.5, 1, 2],
)
\end{minted}
The bases to make as many of these measurements simultaneously as possible are chosen. For this example, this is simply the computational basis. 
Currently, for $d = 10$ Trotter steps at 1000 samples for three different noise depths, the generation of these circuits takes around an hour on a regular laptop computer. 
The time taken to generate these circuits is currently the biggest bottleneck to the execution time of the protocol. One potential solution would be to take advantage of the consistent form of the PER circuits by using parametric compilation to implement the twirl and partial noise inverse. This will be explored in future versions. 

\begin{figure}[t!]
    \centering 
    \includegraphics[width = \linewidth]{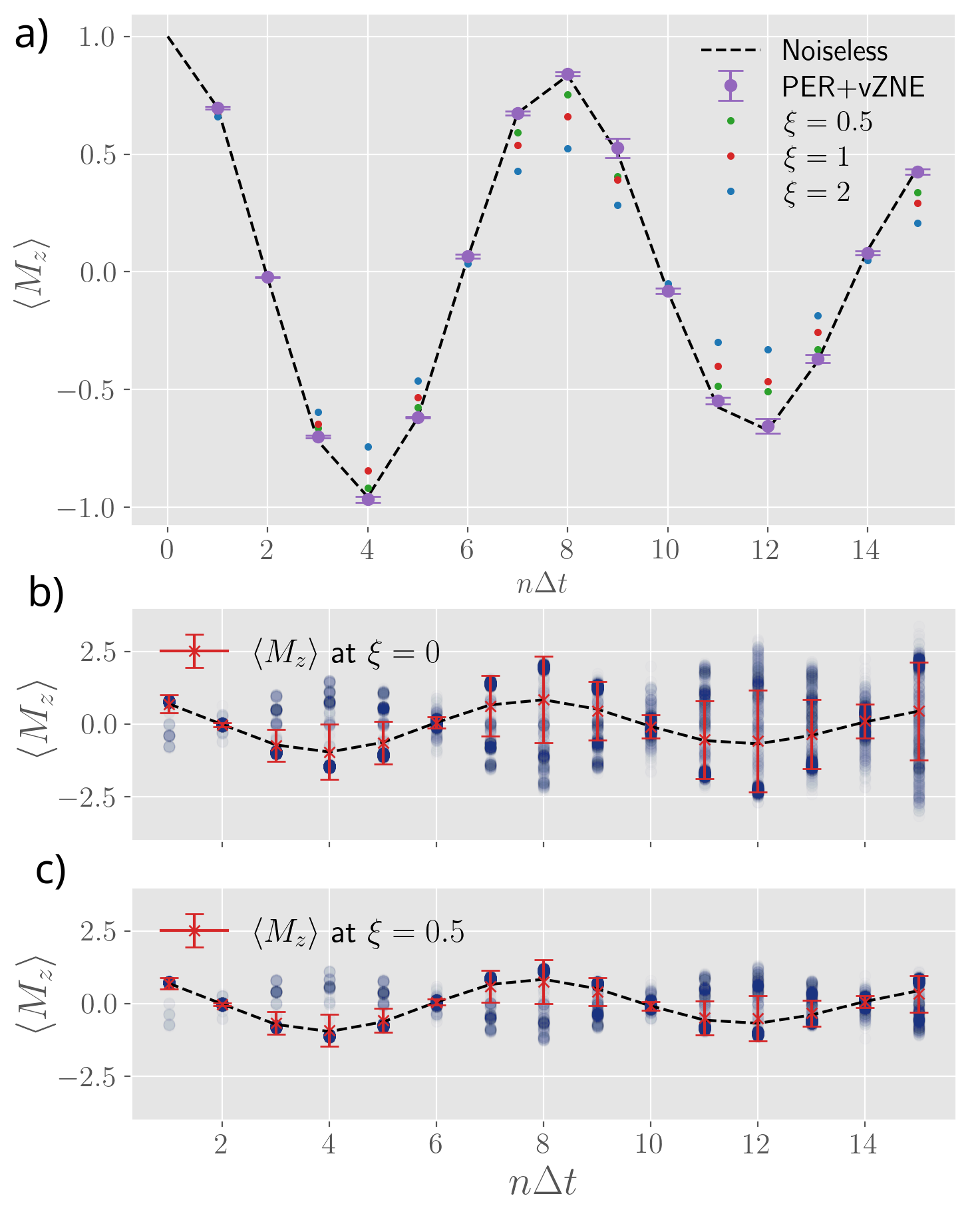}
    \caption{PER results of total magnetization $\langle M_z(t) \rangle $ of the TFIM, prepared in the $|0\rangle$ initial state and evolved with under Hamiltonian with parameters $J = 0.15$, $h = 1$. The Trotter dynamics are simulated on the IBM noisy simulator \texttt{FakeVigoV2} using a Trotter stepsize $\Delta t = 0.2$. We simulate 1000 PER circuits, each is evaluated with $1024$ shots. Panel (a) shows that vZNE with the noise levels $\xi \in [0.5,1,2]$ yields excellent agreement with the noiseless Trotter result. Readout error mitigation is used at all noise levels. Panel (b) shows the individual estimators (blue) and their average with standard deviation (red crosses)  at $\xi = 0$ (upper plot) and $\xi = 0.5$ (lower plot). 
    The variance increases as the noise strength approaches zero, and the estimator values approach the noiseless value. The overhead at $\xi = 0$ after 15 Trotter steps is $\gamma^{(0)} = 7.25$, while it is only $\gamma^{(0.5)} = 2.69$ at $\xi = 0.5$. 
    }
\label{fig:Mz_dynamics_Trotter}
\end{figure}

\begin{figure}
    \centering
    \includegraphics[width = \linewidth]{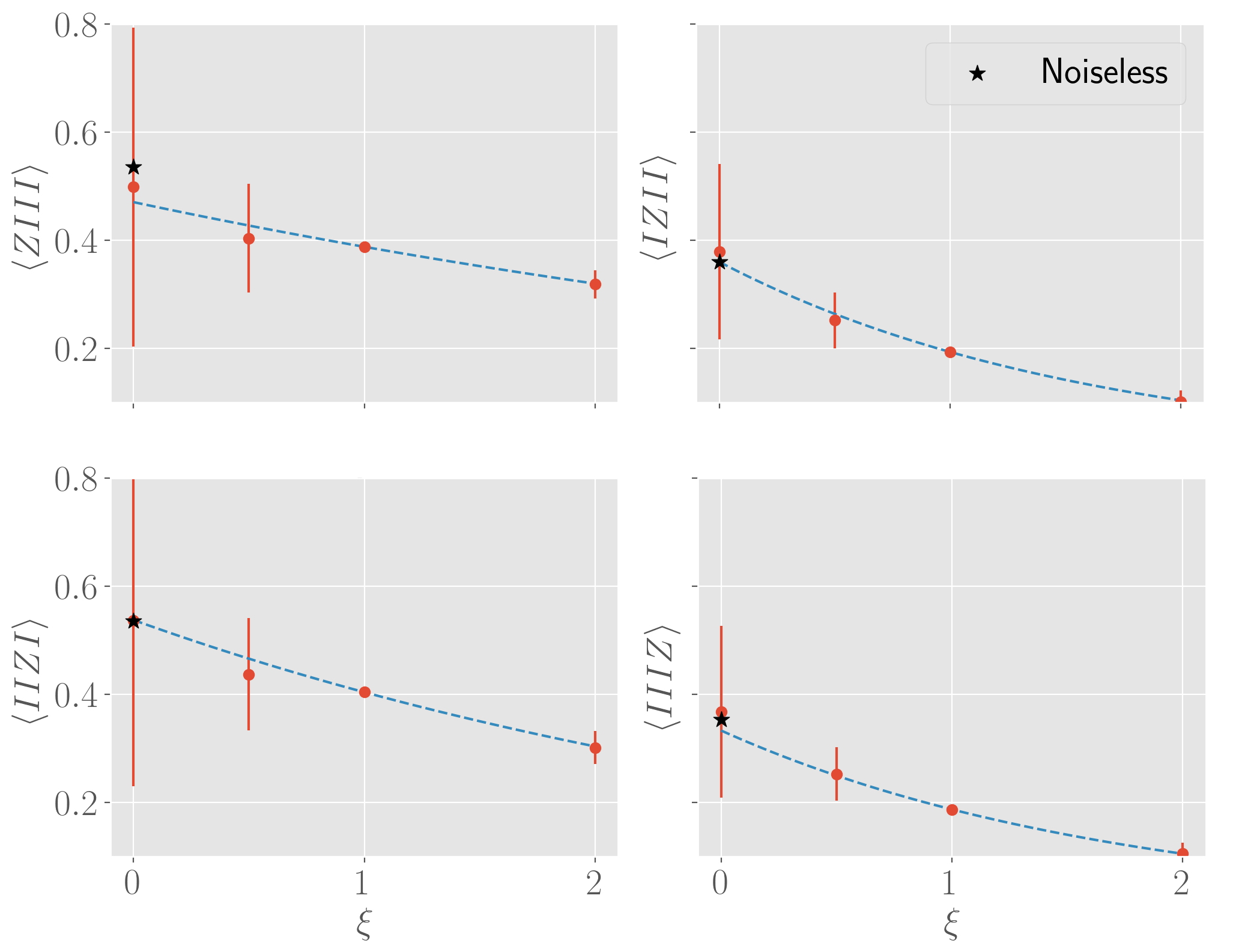}
    \caption{Pauli $Z$ expectation values of individual qubits at $n = 15$ Trotter steps as a function of noise strength $\xi$. We use 1000 PER circuits, each evaluated with 1024 shots. Variance increases and bias decreases for smaller $\xi$. vZNE with exponential fit using $\xi \in \{0.5, 1, 2\}$ yields an expectation value of similar accuracy as PEC. The star denotes the noiseless result. }
    \label{fig:vZNE_Trotter_example}
\end{figure}

After generating the circuits, they can be run with
\begin{minted}{python}
perexp.run(executor)
\end{minted}
Finally, the results can be obtained by calling
\begin{minted}{python}
circuit_results = perexp.analyze()
\end{minted}
Each element in $\code{circuit_results}$ now stores the results of PER containing each expectation value at each noise strength corresponding to the original $\code{circuits}$ array that was passed in. Lastly, the magnetization can be computed for the $i^\text{th}$ Trotter step by adding up the vZNE result of each expectation value:
\begin{minted}{python}
res = circuit_results[i]
M_z[i] = sum([
    res.get_result(op).expectation 
    for op in expectations
]) / N
\end{minted}
The resulting array $M_z$ is plotted versus time in Fig.~\ref{fig:Mz_dynamics_Trotter}(a), along with the noiseless value for comparison. Fig.~\ref{fig:Mz_dynamics_Trotter}(b,c) compares the distribution of PEC and PER estimators, highlighting the increase of the negativity of the QPD for smaller $\xi$ and the resulting larger variance and sampling cost. 
This is further illustrated in Fig.~\ref{fig:vZNE_Trotter_example}, which shows vZNE for the individual $\langle Z_i\rangle$ at the final Trotter step of the simulation. At this depth $n=15$, the number of PEC samples required to achieve a precision equal to $\delta$ is $\frac{{\gamma^{(0)}}^2}{\delta^2} \approx \frac{53}{\delta^2}$. In contrast, PER with the same precision at noise strengths $\xi \in [0.5, 1, 2]$ requires in total only $\frac{{\gamma^{(0.5)}}^2}{\delta^2}+\frac{2}{\delta^2} \approx \frac{9}{\delta^2}$ samples. While there may be some extra bias error introduced through vZNE, our results demonstrate that PER combined with vZNE can offer a significant advantage when the overhead is large.


\section{Conclusion and Outlook}
\label{sec:conclusion}
Mitigating errors that occur in a noisy quantum computation is a key challenge that must be addressed in order to achieve quantum advantage when full quantum error correction is not available. The use of quantum error mitigation is also expected to extend into the era of fault-tolerant quantum computing with logical qubits as it can lower the overhead for quantum error correction~\cite{Lostaglio-PRL-2021,Piveteau-PRL-2021,Suzuki-PRX-2022}. While hardware improvements that reduce gate error rates are crucial, the development of controlled and automated error mitigation software provides progress towards performing more complex quantum calculations. Since the controlled mitigation of quantum errors necessarily includes noise characterization as the first step, we advocate for the development of error mitigation software tools that combine noise characterization with  mitigation capabilities. 

We here provide a Python notebook with a practical example of how to efficiently connect GST with PER using canonical noise scaling and vZNE by connecting two existing software toolkits, PyGSTi and Mitiq. We also develop a software framework that combines PNT with PER using partial noise inversion and vZNE for a user-defined circuit. Since 
the complexity of this method is constant with respect to the number of qubits and the circuit depth for circuits with many repeated layers, this approach can be applied for larger systems. We provide a tutorial notebook showcasing the use of this method in a Trotter simulation of four qubits over 15 Trotter steps. Our results demonstrate that PER combined with vZNE can yield results of the same accuracy as PEC, but at much smaller sampling costs. 
Finally, as increasingly complex error mitigation schemes are developed, multi-platform software tools provide key advantages for software users engaging on multiple quantum computing platforms. We hope that our work lays a foundation for integrating the proposed workflow of GST and canonical noise scaling into Mitiq, and for the development of other software suites connecting noise characterization and error reduction to automate the application of noise-sensitive error mitigation to near-term algorithms on noisy intermediate-scale quantum (NISQ) devices.

\section{Reproducibility}
\label{sec:reproducibility}
We have made our raw data, source code, and tutorial notebooks available via an online appendix~\cite{benjamin_mcdonough_2022_7197234} for the research community to reproduce or use.



\end{document}